\documentclass[%
reprint,
 amsmath,amssymb,
 aps,
prl
]{revtex4-1}

\usepackage{graphicx}
\usepackage{dcolumn}
\usepackage{bm}

\usepackage[usenames]{xcolor}

\begin{document}

\preprint{INT-PUB-18-055}

\title{Gravitational Waves from Compact Dark Objects in Neutron Stars}

\author{C. J. Horowitz}
\email{horowit@indiana.edu}
\affiliation{Center for Exploration of Energy and Matter and Department of Physics, Indiana University, Bloomington, IN 47405, USA}

\author{Sanjay Reddy}
\email{sareddy@uw.edu}
\affiliation{Institute for Nuclear Theory and Department of Physics, University of Washington, Seattle,  WA 98195. }

\date{\today}

\begin{abstract}
Dark matter could be composed of compact dark objects (CDOs).  We find that the oscillation of CDOs inside neutron stars can be a detectable source of gravitational waves (GWs). The GW strain amplitude depends on the mass of the CDO,  and its frequency is typically in the range 3-5 kHz as determined by the central density of the star.  In the best cases, LIGO may be sensitive to CDO masses greater than or of order $10^{-8}\, M_\odot$.  
\end{abstract}

\pacs{Valid PACS appear here}
\maketitle


The distribution of dark matter on very small scales has important implications for the particle nature of dark matter and for primordial  density fluctuations that could produce sub-solar mass black-holes during inflation.  Observing campaigns by the MACHO \cite{Alcock:1995dm,Alcock:2000ph} and EROS \cite{Tisserand:2006zx} collaborations to identify microlensing of stars in the Magellanic clouds by non-luminous compact object in the galactic halo \cite{Paczynski:1986} provide constraints on the population of dark matter objects. Although these constraints rule out scenarios in which dark objects in the mass range $10^{-7}-15$ $M_\odot$ make up most of the dark matter in the milky way halo, they cannot exclude the possibility that a small fraction $\simeq 10\%$ of the mass is contained in a sub-population of dark objects. If such a population exists, interactions between them and neutron stars (NSs) provides a new mechanism to generate gravitational waves (GWs). The purpose of this letter is to show that for dark objects with masses $\gtrsim 10^{-8}$ $M_\odot$ these GWs can be long lived and have an amplitude large enough for detection from nearby NSs in the galaxy. This provides a new way to discover or constrain this population of compact dark objects (CDOs). Intriguingly, we find that such a discovery would allow us to study the interiors of NSs, and the particle properties of dark matter.        

We will entertain the possibility that compact dark objects were formed in the early universe.
Trace amounts of dark matter can also accrete onto NSs from a uniform interstellar medium  \cite{CaptureDM,PhysRevD.40.3221,PhysRevD.82.063531}, but in this case the mass accumulated is too small to be an interesting source of GWs. CDOs produced in the early universe include primordial black-holes (PBHs), and less compact objects made of dark matter particles held together by either gravity or attractive self-interactions in the dark sector. 

 As one concrete example, consider dark matter composed of CDOs each of mass $m_D$ (assumed $\ll M_\odot$).  If the mass density of dark matter near the solar system is $\rho\approx 6\times 10^{-28}$ kg/cm$^3$ \cite{0954-3899-41-6-063101}, the number density of CDOs is $n_{D}=\rho/m_D\approx 3\times 10^{-44}$m$^{-3}\times (10^{-8}M_\odot/m_D)$.  The rate of CDO collisions with a given star is $\sigma\, v_d\, n_{D}$ where $v_d$ is the velocity of the CDO.  For simplicity we consider a single value $v_d\approx 220$ km/s that is adequate for a first rough estimate.  The collision cross section is $\sigma=\pi R^2[1+(v_e/v_d)^2]$ where $R$ is the star's radius and $v_e$ is it's escape velocity.  
 
Let us focus on neutron stars born from massive stars near the galactic center.  Consider $10M_\odot$ stars with lifetimes $T_{10}=2\times 10^7$ y and radii $R_{10}=5.6R_\odot$.  We assume the density of dark matter near the galactic center is a factor of $\rho_{gc}/\rho$ higher than the density near the solar system.  The total number of CDO collisions during the massive star's main sequence lifetime is $N_{\rm coll} = T_{10} \sigma v_d n_{D}( \rho_{gc}/\rho)$,
 \begin{equation}
 N_{\rm coll} \approx 5.4\times \Bigl(\frac{10^{-8}M_\odot}{m_D}\Bigr)\Bigl(\frac{\rho_{gc}/\rho}{1000}\Bigr)\, .
 \end{equation}
If $m_D$ is $10^{-8}M_\odot$ or less, essentially every such star will suffer collisions with CDOs.  Furthermore, if the unknown non-gravitational interactions are strong enough to capture a significant fraction of the CDOs that do collide, most of these stars should contain one or more CDO (by the time they explode as a SN).   Note that $N_{\rm coll}$ is somewhat larger for $1M_\odot$ stars that live for $10^{10}$y.  In this paper we explore GW radiation from CDOs {\it in} or near neutron stars.


The LIGO-Virgo collaboration recently searched for GWs from binaries with sub-solar mass compact objects with masses in the range $0.2-1~M_\odot$ and set useful limits \cite{Abbott:2018oah}. Gravitational waves from the merger of dark boson stars were calculated in ref. \cite{Bezares:2018qwa}.  The rates for GWs from a population of binary black-holes over wider mass-range formed in ``atomic dark matter'' scenarios has also been studied recently \cite{Shandera:2018xkn}. Here structure formation in the dark sector proceeds by processes similar to those that occur in hydrogen clouds and can produce CDOs, both compact dark stars and black-holes with wide-ranging masses\cite{Buckley:2017ttd}. Ellis et al. have suggested that dark matter in NSs could lead to a noticeable modification of the GW signal following a conventional NS merger \cite{Ellis:2017jgp}.  However this may require 0.05 $M_\odot$ or more of dark matter in the NSs.  In addition to GWs, LIGO may also be sensitive to new long range interactions with dark matter, for example from a dark photon field \cite{DarkphotonDM}, or a light scalar  \cite{Grabowska:2018lnd}.

Stellar core collapse is expected to require at least a Chandrashekar mass \cite{Suwa2018}.  Indeed the lightest, well measured, NS has a mass of $1.174\pm 0.004\, M_\odot$ \cite{LowmassNS}.  However GW observation of a compact object with a mass between 0.1 and 1 $M_\odot$ doesn't necessarily prove the existence of exotic new forms of matter.  The  object could still be a NS that was formed in an unexpected way.  

In contrast, the observation of even a single GW event involving a compact object with mass below $0.1M_\odot$ has immediate and extraordinary implications.  It would prove the existence of a new kind of exotic compact object such as a primordial black hole, an object of dark matter, or a star made of self-bound QCD (strange) matter.    Neutron matter is not bound.  It take a minimum NS mass before gravity can overcome the neutron matter repulsion and bind the star. The minimum mass for a NS is around $0.1M_\odot$.  A NS below this mass is unbound and would explode \cite{ExplodingNS}. Therefore, in this letter we focus on GWs from compact objects with masses below $0.1M_\odot$ orbiting near, {\it or in}, a conventional NS.  The discovery of such a low mass object would have dramatic implications. In what follows we shall assume that CDOs are much smaller than NSs, and neglect interesting effects that can arise due to their deformability that could provide information about the nature of dark matter \cite{Nelson:2018xtr}.

The frequency of the oscillation or orbit of a dark matter object deep inside a NS is 
\begin{equation}
    \nu_c=\frac{1}{2\pi} \sqrt{\frac{4\pi G \rho_c}{3}}= 841  \sqrt{\frac{\rho_c}{10^{14} {\rm g/cm^3}}}   ~{\rm Hz} \,. 
\label{eq:nu_c}
\end{equation}
This is set by the central density $\rho_c$ of the compact object, for NSs with $\rho_c \simeq 10^{15}$ g/cm$^3$, the gravitational wave frequency $f_{GW} =2 \nu_c \simeq 5.3\,$kHz. 
More generally, if dark objects are captured from the interstellar medium they will orbit or oscillate in the gravitational potential of the compact object of mass $M$ and radius $R$.

We now calculate the orbital lifetime $T$ of a dark matter object against GW radiation.  We assume the mass of the CDO $m_D$ is much less than the mass of the star and the velocity of the object is much less than $c$. The luminosity due to quadrupole GW radiation is,
\begin{equation}
\frac{dE}{dt}=  \frac{G}{5c^5}\sum_{i,j}\Bigl(\frac{d^3}{dt^3}I_{ij}^T\Bigr)^2\, ,
\end{equation}
where $I_{ij}^T$ is the traceless quadrupole moment tensor \cite{Poisson_Will}.

For simplicity we consider a circular orbit of the object of radius $r$ (that is less than the radius of the star) and frequency $\omega=2\pi \nu$.  We have 
\begin{equation}
\frac{1}{T}=\frac{1}{E}\frac{dE}{dt}=\frac{32 G}{5 c^5}m_D r^2 \omega^4\, .    
\end{equation}
Here $T$ is a characteristic lifetime and 
\begin{equation}
  E=m_Dr^2\omega^2\,,  
  \label{Eq.E}
\end{equation}
is the energy of the CDO measured relative to its potential energy at the center of the star.  
The GW orbital lifetime $T$ of dark matter in a NS is of order,
\begin{equation}
    T\approx {\rm 68\,  h}\, \bigl(\frac{M_\odot\, {\rm m}^2}{m_D\, r^2}\bigr)\bigl(\frac{3.3 {\rm kHz}}{f_{GW}}\bigr)^4\, . 
    \label{Eq.TNS}
\end{equation}
Note that this lifetime only considers GW radiation.  Additional dissipation from, for example, bulk or shear viscosity could reduce $T$.

The gravitational wave strain at earth $h_{jk}$ from the dark matter motion is,
\begin{equation}
    h_{jk}=\frac{2G}{c^4d}\frac{d^2}{dt^2}I_{jk}^{TT}\, ,
\end{equation}
here $d$ is the distance to the source and $I_{jk}^{TT}$ is the transverse traceless quadrupole moment \cite{Poisson_Will}.  A characteristic strain amplitude $h_0=h_{xx}-h_{yy}$ is,
\begin{equation}
    h_0\approx \frac{4G}{c^4 d} m_Dr^2 \omega^2\, ,
    \label{Eq.h0}
\end{equation}
where we assume the orbital plane is perpendicular to the observing direction.

The strain depends on the product of dark matter object mass times radius squared $m_Dr^2$.  As an example we consider an object in a NS $d=1$ kpc away with $m_Dr^2$ of order 1 $M_\odot\, m^2$ orbiting at a gravitational wave frequency $f_{GW}=\omega/\pi=5.3$ kHz,
\begin{equation}
    h_0\approx 1.4\times 10^{-25} \bigl(\frac{m_D\, r^2}{M_\odot\, m^2}\bigr)\bigl(\frac{1\, kpc}{d}\bigr)\bigl(\frac{f_{GW}}{5.3\, \rm kHz}\bigr)^2\, .
    \label{Eq.h0NS}
\end{equation}


We now explore the detectability of this GW signal in LIGO.  
We start by considering large amplitude motion where the CDO, originally orbiting outside the star, merges because of GW radiation.  Outside the star we have a conventional inspiral ``chirp'' that is characterized by a chirp mass $M_c=\mu^{3/5}M_{tot}^{2/5}$,
where $M_{tot}$ is the total mass of the dark object plus the NS and $\mu$ is the reduced mass.  If $m_D\ll M_{NS}$ then the chirp mass $M_c\approx m_D^{3/5}M_{NS}^{2/5}$ is small.

\begin{figure}[ht]
\smallskip
\includegraphics[width=1.\columnwidth]{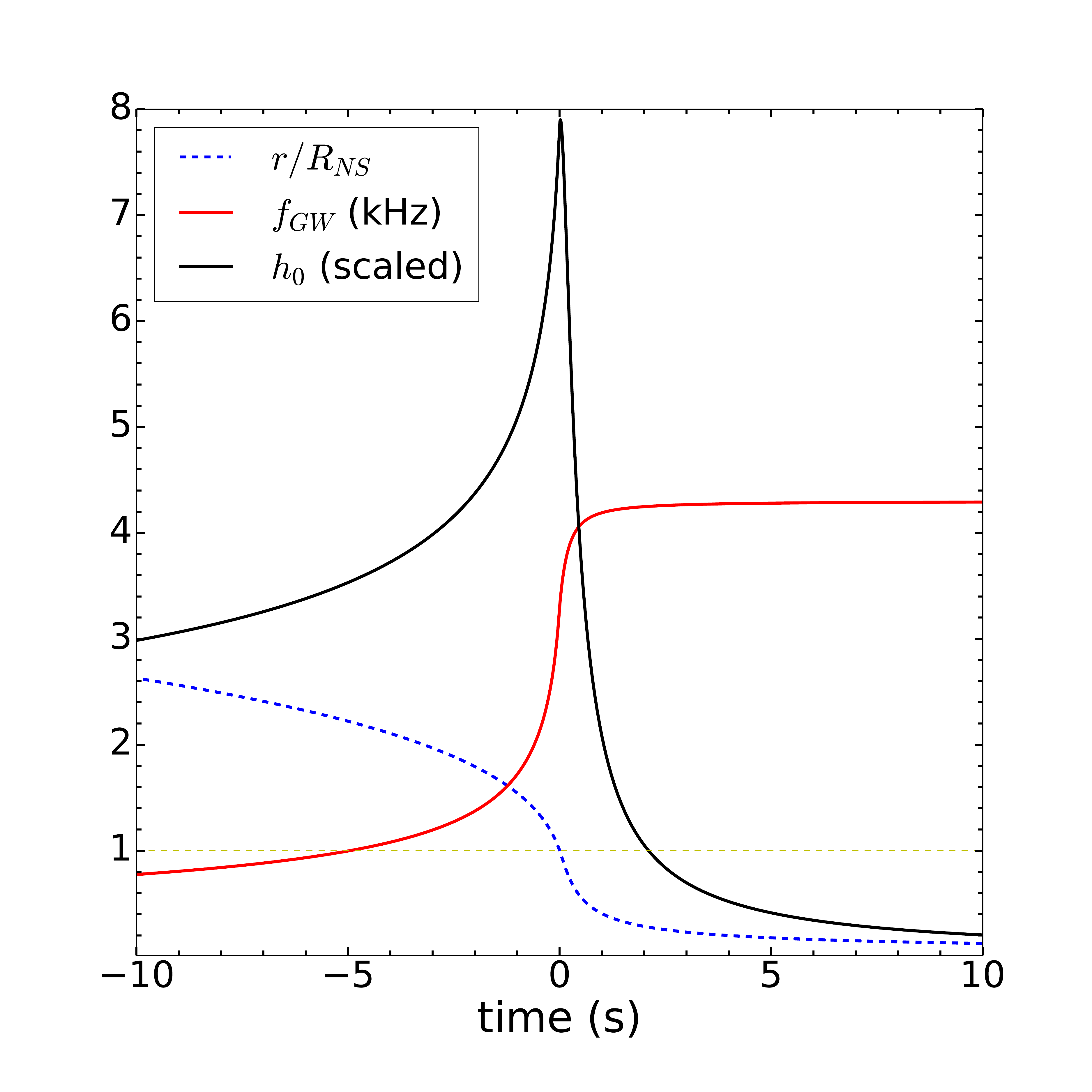}
 \caption{(Color online) 
Orbital radius $r$ in units of the NS radius $R_{NS}$ (blue dashed curve), gravitational wave frequency $f_{GW}$ (red curve, in kHz), and GW strain $h_0$ (solid black curve, arbitrarily scaled) for a $m_D=10^{-3}M_\odot$ dark matter object in a circular orbit around, and then in, a NS of radius 12 km and mass $1.4~M_\odot$ as a function time $t$. For  $t>0$ the orbit of the dark object resides inside the NS.}
\label{Fig1}
\end{figure}

The conventional chirp ends when the orbital radius equals the star radius and the dark object enters the star.  We assume that the dark matter interacts very weakly with conventional matter.  As a result the dark matter object can continue to move inside the star and will radiate GW according to Eq.~\ref{Eq.h0}.  Thus the GW signal will have both an outside part and then an inside part where the dark matter is orbiting inside the star.  We refer to this as a ``common envelope chirp".  In Fig.~\ref{Fig1} we show the radius $r$, GW frequency $f_{GW}$ and strain amplitude $h_0$ (arbitrarily scaled) vs time for a dark object with mass $m_D=10^{-3} M_\odot$ orbiting a 1.4$M_\odot$ NS.  The NS radius is $R_{NS}=12$ km and we choose a simple analytic density profile $\rho(r)=\rho_c(1-(r/R_{NS})^2)^{1/2}$ that mimics profiles obtained using realistic dense matter equations of state.

For times $t<0$, in Fig.~\ref{Fig1}, the object is outside the NS and one has a normal chirp where $f_{GW}$ and $h_0$ increase with time.  Then when $r=R_{NS}$ at $t=0$ the character of the evolution changes so that $h_0$ starts to decrease and the evolution of $f_{GW}$ with time depends on the density profile of the NS. If the density inside the NS was assumed to be constant, $f_{GW}=2\nu_c$ would also be constant. As noted earlier, we have chosen a simple but realistic ansatz for the density profile, and this is reflected in the time dependence of $f_{GW}$.  
The maximum value of $h_0$ in Fig.~\ref{Fig1} occurs when $r=R_{NS}$ and is given by Eq.~\ref{Eq.h0}
\begin{equation}
    h_0^{max}=3.3\times 10^{-17} \bigl(\frac{m_D}{M_\odot}\bigr)\bigl(\frac{R_{NS}}{12\ {\rm km}}\bigr)^2\bigl(\frac{f_{GW}}{3.3\, {\rm kHz}}\bigr)^2\bigl(\frac{{\rm kpc}}{d}\bigr).
    \label{Eq.h0max}
\end{equation}
The characteristic lifetime $T^*$ is given by Eq.~\ref{Eq.TNS} with $r=R_{NS}$,
\begin{equation}
    T^*=1.7\, {\rm ms}\,  \bigl(\frac{M_\odot}{m_D}\bigr)\bigl(\frac{12\, {\rm km}}{R_{NS}}\bigr)^2\bigl(\frac{3.3\, {\rm kHz}}{f_{GW}}\bigr)^4\, .
\end{equation}
For Fig.~\ref{Fig1} with $m_D=10^{-3}\ M_\odot$ the peak in $h_0$ has a width $T^*\approx 1.7$\, s.

We emphasize the extraordinary information that might be contained in a single observation similar to Fig.~\ref{Fig1}.   The small chirp mass by itself, if less than the $\approx 0.1M_\odot$ minimum mass of a NS, implies the presence of an exotic compact object as NSs below this mass are not bound. Possibilities include self bound QCD matter (presumably strange matter), primordial black holes, or an object of dark matter. Self-bound QCD matter will interact strongly with the NS and lead to an $h_0$ that, after the peak, is likely much smaller than what is shown in Fig.~\ref{Fig1}.  
It may be possible to distinguish between CDO that are primordial black holes (PBH) and those that are not because, as we discuss later, a PBH that merges with a NS will accrete matter and eventually cause the NS to collapse, and this collapse may be observable in GW.  Alternatively, the merger of a PBH with a stellar mass BH will have a GW signal that fades away more rapidly after the merger than the signal from a non-BH CDO merging with a NS.  Finally, the post merger frequency of the GW signal provides direct information about the density profile of the NS (and thus the equation of state).


The detectability of a nearly continuous GW signal may depend primarily upon the amplitude $h_0^{max}$ and somewhat on the observation time $T_{obs}$.   If one could coherently integrate the signal over a full time $T_{obs}$ the signal to noise ratio might improve with $T_{obs}^{1/2}$.  In practise this could be computationally too expensive and a computationally easier, but somewhat less sensitive search would be employed.

For simplicity, we discuss the detectability of a signal in terms of the sensitivity depth $D$ of a search that is necessary for detection \cite{Behnke:2014tma,Walsh:2016hyc,Dreissigacker:2018afk}.  The sensitivity depth $D$ is defined as the square root of the noise amplitude spectral density $\sqrt{S_n(f_{GW})}$ divided by the strain amplitude $h_0$ (see Eq. 22 of ref. \cite{Behnke:2014tma}),
\begin{equation}
    D(f_{GW})=\frac{\sqrt{S_n(f_{GW})}}{h_0}\, .
\end{equation}
The lower the value of $D$ the easier it is to find a signal.  Advanced LIGO, at target sensitivity, may have $\sqrt{S_n(3\ {\rm kHz})}\approx 3\times 10^{-23}$ Hz$^{-1/2}$ \cite{Abbott2018,PhysRevD.93.112004}. Note that LIGO has not yet reached this sensitivity but increasing the laser power and the use of squeezed light \cite{Squeeze} should reduce $\sqrt{S_n(f)}$ at these high frequencies.  Using this sensitivity and $h_0^{max}$ from Eq.~\ref{Eq.h0max} we have,
\begin{equation}
 D(3.3\,{\rm kHz})\approx 9.1\times 10^{-7}\ {\rm Hz}^{-1/2}\Bigl(\frac{M_\odot}{m_D}\Bigr)\Bigl(\frac{d}{\rm kpc}\Bigr)\, .  
\end{equation}
As a rough guide, a signal could likely be found if $D<50$ Hz$^{-1/2}$\, \cite{Dreissigacker:2018afk}.  This suggests a sensitivity to CDO masses larger than,
\begin{equation}
    m_D>1.8\times 10^{-8}M_\odot\, , 
\end{equation}
at a distance of 1 kpc or $m_D>1.8\times 10^{-7}M_\odot$ at 10 kpc.  Thus LIGO is potentially sensitive to very small CDO masses much less than a solar mass.  

We now consider small amplitude motion of dark matter objects in a NS with $r\ll R_{NS}$.  For example, a dark matter object originally at the center of a star is excited so that it undergoes small amplitude oscillations about $r=0$.  Note that $f_{GW}$ depends slightly on $r$.  For the example in Fig. \ref{Fig1}, $f_{GW}$ increases by about 25\% as $r$ decreases from $R_{NS}$ to $0$. 
The amplitude of the GW radiation depends on the combination $m_D r^2$.  It is convenient to parameterize this in terms of the energy $E$ of a circular orbit of radius $r$, see Eq.~\ref{Eq.E}.  Results for eccentric orbits are expected to be similar.  In terms of $E$ the GW amplitude is,
$h_0\approx 4GE/(c^4 d)$, and the lifetime against GW radiation is $T=5c^5/(32G\omega^2E)$.  The sensitivity depth can now be written,
$D\approx ({2.7\times 10^{47}\, {\rm ergs}}/{E})({d}/{\rm kpc}){\rm Hz}^{-1/2}$.
This suggests LIGO could detect ($D<50$ Hz$^{-1/2}$) CDOs excited to energies larger than,
\begin{equation}
    E\ge 5.4\times 10^{45}\, {\rm ergs}\, \Bigl(\frac{d}{\rm kpc}\Bigr)\, .
    \label{Eq.Eneeded}
\end{equation}
Core collapse SN with kinetic energies of order $10^{51}$ ergs have plenty of energy to excite a CDO. Observed NS kicks with velocities of several hundred km/s suggest that a fraction of this kinetic energy is imparted aspherically \cite{Holland-Ashford:2017tqz} and is adequate to excite CDOs. Thus a galactic SN could emit both a short burst of GW during the explosion and then a nearly continuous high frequency GW signal for days afterwards.  A LIGO search for continuous GW from the direction of SN1987A may have occurred too late after the SN \cite{PhysRevLett.107.271102}. 

As discussed in the introduction, one or more CDO may be present in a pre-SN star.  What happens to them during the explosion?  If an object is originally at the star's center and both the collapse and explosion are spherically symmetric, the CDO could remain at the center and not be excited.  However, this may be unlikely.  Violent convection in the burning Si/O shells is expected to produce turbulent fluctuations during the final pre-SN evolution \cite{1994ApJ...427..932A,2011ApJ...733...78A,2041-8205-808-1-L21}.  These fluctuations could excite the CDO before the core collapses.  In addition, many SN explosions are asymmetric and produce large recoil velocities for the newly formed NS \cite{2041-8205-725-1-L106,refId0}.  These asymmetries could further excite the CDO during the SN explosion.  This could lead to a CDO that orbits inside the NS or is even temporarily ejected to fallback or inspiral at a later time.  The excitation energy of CDOs should be explored in, possibly simplified, SN simulations.  

Magnetar giant flares with energies of order $10^{46}$ ergs \cite{MagnetarGF} may also be able to excite embedded CDOs.  The high frequency GW emission from a CDO in a magnetar need not be constrained by the low rotational frequency of the magnetar or by its very rapid electromagnetic spin down.  There have been several searches for bursts of GW during giant flares, see for example \cite{PhysRevD.85.024030,LIGO_SGR1806}.  We suggest instead to search for high frequency, nearly continuous, GW radiation that may last for days after the flare.

If CDOs in compact stars are low mass primordial black holes (PBHs), they may not be captured by main sequence stars because they have no non-gravitational interactions \cite{0004-637X-705-1-659}.  Instead direct PBH-NS collisions may be necessary.  These are rare because NS are small.  We estimate at best a total rate of NS-PBH collisions in the Galaxy of $\approx 10^{-5} (10^{-8}M_\odot/m_D)$\, y$^{-1}$.  For $m_D\approx 10^{-12}M_\odot$ the rate could be high enough.  However, in this case GW from the inspiral may be too weak for LIGO to detect.  Instead the PBH will accrete matter and cause the NS to collapse in less than a year \cite{0004-637X-868-1-17}.  This collapse will radiate GW with an amplitude estimated to be $h_0\approx 5\times 10^{-28}({\rm 1kpc}/d)(1{\rm s}/T)^2$ where $d$ is the distance to the NS and $T$ is the collapse time scale  \cite{0004-637X-868-1-17}.  For $T$ equal to the ms dynamical time scale of a NS, LIGO likely could detect this signal in our Galaxy.  Note  Abramowitz et al suggest such a collapse could also produce a fast radio burst \cite{0004-637X-868-1-17}.

There have been several previous LIGO searches for both burst and continuous GW signals.  For example searches in the direction of the galactic center are promising because this could be a region of high dark matter density \cite{Behnke:2014tma,PhysRevD.88.102002}.

We find strong GW can be emitted by even a very low mass CDO orbiting (in) a NS.  In contrast, two low mass CDOs orbiting each other may emit a much weaker GW signal because the amplitude $r$ in Eq.~\ref{Eq.h0} is constrained to small values in order for the objects to orbit at a high enough frequency.

Detailed modeling of the expected GW signal may be necessary for the most sensitive searches.  In particular, a detailed understanding of the evolution of $f_{GW}$ with time may be important.  To zeroth order we find $f_{GW}$ to be independent of time, for small amplitude motion.  However there could be small corrections.  We assumed the density distribution of the NS is frozen.  In reality it will respond to the CDO motions.  This could change $f_{GW}$ and lead to additional dissipation.   

To summarize, we have proposed a novel source of detectable GWs which might reveal a new population of compact dark objects (CDOs). Should they exist, despite our poor understanding of mechanisms for CDO formation and their capture by NSs, our proposal allows for their detection when their masses $\gtrsim 10^{-8}~M_\odot$. Their discovery would have far reaching implications for dark matter and provide a means to directly probe the interiors of NSs. The detectable signatures we discuss are robust, but additional work may be needed to produce the best templates for GW searches. 
If the dark objects are not compact their tidal deformabilities could be large and will alter the GW emission with potentially observable signatures \cite{Nelson:2018xtr}. 

We thank Maria Alessandra Papa for helpful comments.  CJH thanks the Institute for Nuclear Theory for its hospitality.  CJH is supported in part by DOE grants DE-FG02-87ER40365 and DE-SC0018083.  SR acknowledges support from the US Department of Energy Grant No. DE-FG02-00ER41132.  C. J. H.and S. R. also acknowledge support from  the National Science Foundation Grant No. PHY-1430152 (JINA Center for the Evolution of the Elements).
\bibliographystyle{apsrev}

\begin{thebibliography}{39}
\expandafter\ifx\csname natexlab\endcsname\relax\def\natexlab#1{#1}\fi
\expandafter\ifx\csname bibnamefont\endcsname\relax
  \def\bibnamefont#1{#1}\fi
\expandafter\ifx\csname bibfnamefont\endcsname\relax
  \def\bibfnamefont#1{#1}\fi
\expandafter\ifx\csname citenamefont\endcsname\relax
  \def\citenamefont#1{#1}\fi
\expandafter\ifx\csname url\endcsname\relax
  \def\url#1{\texttt{#1}}\fi
\expandafter\ifx\csname urlprefix\endcsname\relax\def\urlprefix{URL }\fi
\providecommand{\bibinfo}[2]{#2}
\providecommand{\eprint}[2][]{\url{#2}}

\bibitem[{\citenamefont{Alcock et~al.}(1995)}]{Alcock:1995dm}
\bibinfo{author}{\bibfnamefont{C.}~\bibnamefont{Alcock}} \bibnamefont{et~al.}
  (\bibinfo{collaboration}{MACHO}), \bibinfo{journal}{Phys. Rev. Lett.}
  \textbf{\bibinfo{volume}{74}}, \bibinfo{pages}{2867} (\bibinfo{year}{1995}),
  \eprint{astro-ph/9501091}.

\bibitem[{\citenamefont{Alcock et~al.}(2000)}]{Alcock:2000ph}
\bibinfo{author}{\bibfnamefont{C.}~\bibnamefont{Alcock}} \bibnamefont{et~al.}
  (\bibinfo{collaboration}{MACHO}), \bibinfo{journal}{Astrophys. J.}
  \textbf{\bibinfo{volume}{542}}, \bibinfo{pages}{281} (\bibinfo{year}{2000}),
  \eprint{astro-ph/0001272}.

\bibitem[{\citenamefont{Tisserand et~al.}(2007)}]{Tisserand:2006zx}
\bibinfo{author}{\bibfnamefont{P.}~\bibnamefont{Tisserand}}
  \bibnamefont{et~al.} (\bibinfo{collaboration}{EROS-2}),
  \bibinfo{journal}{Astron. Astrophys.} \textbf{\bibinfo{volume}{469}},
  \bibinfo{pages}{387} (\bibinfo{year}{2007}), \eprint{astro-ph/0607207}.

\bibitem[{\citenamefont{{Paczynski}}(1986)}]{Paczynski:1986}
\bibinfo{author}{\bibfnamefont{B.}~\bibnamefont{{Paczynski}}},
  \bibinfo{journal}{\apj} \textbf{\bibinfo{volume}{304}}, \bibinfo{pages}{1}
  (\bibinfo{year}{1986}).

\bibitem[{\citenamefont{GÃŒver et~al.}(2014)\citenamefont{GÃŒver, Erkoca,
  Reno, and Sarcevic}}]{CaptureDM}
\bibinfo{author}{\bibfnamefont{T.}~\bibnamefont{GÃŒver}},
  \bibinfo{author}{\bibfnamefont{A.~E.} \bibnamefont{Erkoca}},
  \bibinfo{author}{\bibfnamefont{M.~H.} \bibnamefont{Reno}}, \bibnamefont{and}
  \bibinfo{author}{\bibfnamefont{I.}~\bibnamefont{Sarcevic}},
  \bibinfo{journal}{Journal of Cosmology and Astroparticle Physics}
  \textbf{\bibinfo{volume}{2014}}, \bibinfo{pages}{013} (\bibinfo{year}{2014}),
  \urlprefix\url{http://stacks.iop.org/1475-7516/2014/i=05/a=013}.

\bibitem[{\citenamefont{Goldman and Nussinov}(1989)}]{PhysRevD.40.3221}
\bibinfo{author}{\bibfnamefont{I.}~\bibnamefont{Goldman}} \bibnamefont{and}
  \bibinfo{author}{\bibfnamefont{S.}~\bibnamefont{Nussinov}},
  \bibinfo{journal}{Phys. Rev. D} \textbf{\bibinfo{volume}{40}},
  \bibinfo{pages}{3221} (\bibinfo{year}{1989}),
  \urlprefix\url{https://link.aps.org/doi/10.1103/PhysRevD.40.3221}.

\bibitem[{\citenamefont{Kouvaris and Tinyakov}(2010)}]{PhysRevD.82.063531}
\bibinfo{author}{\bibfnamefont{C.}~\bibnamefont{Kouvaris}} \bibnamefont{and}
  \bibinfo{author}{\bibfnamefont{P.}~\bibnamefont{Tinyakov}},
  \bibinfo{journal}{Phys. Rev. D} \textbf{\bibinfo{volume}{82}},
  \bibinfo{pages}{063531} (\bibinfo{year}{2010}),
  \urlprefix\url{https://link.aps.org/doi/10.1103/PhysRevD.82.063531}.

\bibitem[{\citenamefont{Read}(2014)}]{0954-3899-41-6-063101}
\bibinfo{author}{\bibfnamefont{J.~I.} \bibnamefont{Read}},
  \bibinfo{journal}{Journal of Physics G: Nuclear and Particle Physics}
  \textbf{\bibinfo{volume}{41}}, \bibinfo{pages}{063101}
  (\bibinfo{year}{2014}),
  \urlprefix\url{http://stacks.iop.org/0954-3899/41/i=6/a=063101}.

\bibitem[{\citenamefont{Abbott et~al.}(2018{\natexlab{a}})}]{Abbott:2018oah}
\bibinfo{author}{\bibfnamefont{B.~P.} \bibnamefont{Abbott}}
  \bibnamefont{et~al.} (\bibinfo{collaboration}{Virgo, LIGO Scientific})
  (\bibinfo{year}{2018}{\natexlab{a}}), \eprint{1808.04771}.

\bibitem[{\citenamefont{Bezares and Palenzuela}(2018)}]{Bezares:2018qwa}
\bibinfo{author}{\bibfnamefont{M.}~\bibnamefont{Bezares}} \bibnamefont{and}
  \bibinfo{author}{\bibfnamefont{C.}~\bibnamefont{Palenzuela}},
  \bibinfo{journal}{Class. Quant. Grav.} \textbf{\bibinfo{volume}{35}},
  \bibinfo{pages}{234002} (\bibinfo{year}{2018}), \eprint{1808.10732}.

\bibitem[{\citenamefont{Shandera et~al.}(2018)\citenamefont{Shandera, Jeong,
  and Gebhardt}}]{Shandera:2018xkn}
\bibinfo{author}{\bibfnamefont{S.}~\bibnamefont{Shandera}},
  \bibinfo{author}{\bibfnamefont{D.}~\bibnamefont{Jeong}}, \bibnamefont{and}
  \bibinfo{author}{\bibfnamefont{H.~S.~G.} \bibnamefont{Gebhardt}},
  \bibinfo{journal}{Phys. Rev. Lett.} \textbf{\bibinfo{volume}{120}},
  \bibinfo{pages}{241102} (\bibinfo{year}{2018}), \eprint{1802.08206}.

\bibitem[{\citenamefont{Buckley and DiFranzo}(2018)}]{Buckley:2017ttd}
\bibinfo{author}{\bibfnamefont{M.~R.} \bibnamefont{Buckley}} \bibnamefont{and}
  \bibinfo{author}{\bibfnamefont{A.}~\bibnamefont{DiFranzo}},
  \bibinfo{journal}{Phys. Rev. Lett.} \textbf{\bibinfo{volume}{120}},
  \bibinfo{pages}{051102} (\bibinfo{year}{2018}), \eprint{1707.03829}.

\bibitem[{\citenamefont{Ellis et~al.}(2018)\citenamefont{Ellis, Hektor,
  H{\"u}tsi, Kannike, Marzola, Raidal, and Vaskonen}}]{Ellis:2017jgp}
\bibinfo{author}{\bibfnamefont{J.}~\bibnamefont{Ellis}},
  \bibinfo{author}{\bibfnamefont{A.}~\bibnamefont{Hektor}},
  \bibinfo{author}{\bibfnamefont{G.}~\bibnamefont{H{\"u}tsi}},
  \bibinfo{author}{\bibfnamefont{K.}~\bibnamefont{Kannike}},
  \bibinfo{author}{\bibfnamefont{L.}~\bibnamefont{Marzola}},
  \bibinfo{author}{\bibfnamefont{M.}~\bibnamefont{Raidal}}, \bibnamefont{and}
  \bibinfo{author}{\bibfnamefont{V.}~\bibnamefont{Vaskonen}},
  \bibinfo{journal}{Phys. Lett.} \textbf{\bibinfo{volume}{B781}},
  \bibinfo{pages}{607} (\bibinfo{year}{2018}), \eprint{1710.05540}.

\bibitem[{\citenamefont{Pierce et~al.}(2018)\citenamefont{Pierce, Riles, and
  Zhao}}]{DarkphotonDM}
\bibinfo{author}{\bibfnamefont{A.}~\bibnamefont{Pierce}},
  \bibinfo{author}{\bibfnamefont{K.}~\bibnamefont{Riles}}, \bibnamefont{and}
  \bibinfo{author}{\bibfnamefont{Y.}~\bibnamefont{Zhao}},
  \bibinfo{journal}{Phys. Rev. Lett.} \textbf{\bibinfo{volume}{121}},
  \bibinfo{pages}{061102} (\bibinfo{year}{2018}),
  \urlprefix\url{https://link.aps.org/doi/10.1103/PhysRevLett.121.061102}.

\bibitem[{\citenamefont{Grabowska et~al.}(2018)\citenamefont{Grabowska, Melia,
  and Rajendran}}]{Grabowska:2018lnd}
\bibinfo{author}{\bibfnamefont{D.~M.} \bibnamefont{Grabowska}},
  \bibinfo{author}{\bibfnamefont{T.}~\bibnamefont{Melia}}, \bibnamefont{and}
  \bibinfo{author}{\bibfnamefont{S.}~\bibnamefont{Rajendran}}
  (\bibinfo{year}{2018}), \eprint{1807.03788}.

\bibitem[{\citenamefont{Suwa et~al.}(2018)\citenamefont{Suwa, Yoshida, Shibata,
  Umeda, and Takahashi}}]{Suwa2018}
\bibinfo{author}{\bibfnamefont{Y.}~\bibnamefont{Suwa}},
  \bibinfo{author}{\bibfnamefont{T.}~\bibnamefont{Yoshida}},
  \bibinfo{author}{\bibfnamefont{M.}~\bibnamefont{Shibata}},
  \bibinfo{author}{\bibfnamefont{H.}~\bibnamefont{Umeda}}, \bibnamefont{and}
  \bibinfo{author}{\bibfnamefont{K.}~\bibnamefont{Takahashi}},
  \bibinfo{journal}{Monthly Notices of the Royal Astronomical Society}
  \textbf{\bibinfo{volume}{481}}, \bibinfo{pages}{3305} (\bibinfo{year}{2018}).

\bibitem[{\citenamefont{Martinez et~al.}(2015)\citenamefont{Martinez, Stovall,
  Freire, Deneva, Jenet, McLaughlin, Bagchi, Bates, and Ridolfi}}]{LowmassNS}
\bibinfo{author}{\bibfnamefont{J.~G.} \bibnamefont{Martinez}},
  \bibinfo{author}{\bibfnamefont{K.}~\bibnamefont{Stovall}},
  \bibinfo{author}{\bibfnamefont{P.~C.~C.} \bibnamefont{Freire}},
  \bibinfo{author}{\bibfnamefont{J.~S.} \bibnamefont{Deneva}},
  \bibinfo{author}{\bibfnamefont{F.~A.} \bibnamefont{Jenet}},
  \bibinfo{author}{\bibfnamefont{M.~A.} \bibnamefont{McLaughlin}},
  \bibinfo{author}{\bibfnamefont{M.}~\bibnamefont{Bagchi}},
  \bibinfo{author}{\bibfnamefont{S.~D.} \bibnamefont{Bates}}, \bibnamefont{and}
  \bibinfo{author}{\bibfnamefont{A.}~\bibnamefont{Ridolfi}},
  \bibinfo{journal}{The Astrophysical Journal} \textbf{\bibinfo{volume}{812}},
  \bibinfo{pages}{143} (\bibinfo{year}{2015}),
  \urlprefix\url{http://stacks.iop.org/0004-637X/812/i=2/a=143}.

\bibitem[{\citenamefont{{Colpi} et~al.}(1989)\citenamefont{{Colpi}, {Shapiro},
  and {Teukolsky}}}]{ExplodingNS}
\bibinfo{author}{\bibfnamefont{M.}~\bibnamefont{{Colpi}}},
  \bibinfo{author}{\bibfnamefont{S.~L.} \bibnamefont{{Shapiro}}},
  \bibnamefont{and} \bibinfo{author}{\bibfnamefont{S.~A.}
  \bibnamefont{{Teukolsky}}}, \bibinfo{journal}{\apj}
  \textbf{\bibinfo{volume}{339}}, \bibinfo{pages}{318} (\bibinfo{year}{1989}).

\bibitem[{\citenamefont{Nelson et~al.}(2018)\citenamefont{Nelson, Reddy, and
  Zhou}}]{Nelson:2018xtr}
\bibinfo{author}{\bibfnamefont{A.}~\bibnamefont{Nelson}},
  \bibinfo{author}{\bibfnamefont{S.}~\bibnamefont{Reddy}}, \bibnamefont{and}
  \bibinfo{author}{\bibfnamefont{D.}~\bibnamefont{Zhou}}
  (\bibinfo{year}{2018}), \eprint{1803.03266}.

\bibitem[{\citenamefont{Poisson and Will}(2014)}]{Poisson_Will}
\bibinfo{author}{\bibfnamefont{E.}~\bibnamefont{Poisson}} \bibnamefont{and}
  \bibinfo{author}{\bibfnamefont{C.~M.} \bibnamefont{Will}},
  \emph{\bibinfo{title}{Gravity: Newtonian, Post-Newtonian, Relativistic}}
  (\bibinfo{publisher}{Cambridge University Press}, \bibinfo{year}{2014}).

\bibitem[{\citenamefont{Behnke et~al.}(2015)\citenamefont{Behnke, Papa, and
  Prix}}]{Behnke:2014tma}
\bibinfo{author}{\bibfnamefont{B.}~\bibnamefont{Behnke}},
  \bibinfo{author}{\bibfnamefont{M.~A.} \bibnamefont{Papa}}, \bibnamefont{and}
  \bibinfo{author}{\bibfnamefont{R.}~\bibnamefont{Prix}},
  \bibinfo{journal}{Phys. Rev.} \textbf{\bibinfo{volume}{D91}},
  \bibinfo{pages}{064007} (\bibinfo{year}{2015}), \eprint{1410.5997}.

\bibitem[{\citenamefont{Walsh et~al.}(2016)}]{Walsh:2016hyc}
\bibinfo{author}{\bibfnamefont{S.}~\bibnamefont{Walsh}} \bibnamefont{et~al.},
  \bibinfo{journal}{Phys. Rev.} \textbf{\bibinfo{volume}{D94}},
  \bibinfo{pages}{124010} (\bibinfo{year}{2016}), \eprint{1606.00660}.

\bibitem[{\citenamefont{Dreissigacker et~al.}(2018)\citenamefont{Dreissigacker,
  Prix, and Wette}}]{Dreissigacker:2018afk}
\bibinfo{author}{\bibfnamefont{C.}~\bibnamefont{Dreissigacker}},
  \bibinfo{author}{\bibfnamefont{R.}~\bibnamefont{Prix}}, \bibnamefont{and}
  \bibinfo{author}{\bibfnamefont{K.}~\bibnamefont{Wette}},
  \bibinfo{journal}{Phys. Rev.} \textbf{\bibinfo{volume}{D98}},
  \bibinfo{pages}{084058} (\bibinfo{year}{2018}), \eprint{1808.02459}.

\bibitem[{\citenamefont{Abbott et~al.}(2018{\natexlab{b}})\citenamefont{Abbott,
  Abbott, Abbott, Abernathy, Acernese, Ackley, Adams, Adams, Addesso, Adhikari
  et~al.}}]{Abbott2018}
\bibinfo{author}{\bibfnamefont{B.~P.} \bibnamefont{Abbott}},
  \bibinfo{author}{\bibfnamefont{R.}~\bibnamefont{Abbott}},
  \bibinfo{author}{\bibfnamefont{T.~D.} \bibnamefont{Abbott}},
  \bibinfo{author}{\bibfnamefont{M.~R.} \bibnamefont{Abernathy}},
  \bibinfo{author}{\bibfnamefont{F.}~\bibnamefont{Acernese}},
  \bibinfo{author}{\bibfnamefont{K.}~\bibnamefont{Ackley}},
  \bibinfo{author}{\bibfnamefont{C.}~\bibnamefont{Adams}},
  \bibinfo{author}{\bibfnamefont{T.}~\bibnamefont{Adams}},
  \bibinfo{author}{\bibfnamefont{P.}~\bibnamefont{Addesso}},
  \bibinfo{author}{\bibfnamefont{R.~X.} \bibnamefont{Adhikari}},
  \bibnamefont{et~al.}, \bibinfo{journal}{Living Reviews in Relativity}
  \textbf{\bibinfo{volume}{21}}, \bibinfo{pages}{3}
  (\bibinfo{year}{2018}{\natexlab{b}}), ISSN \bibinfo{issn}{1433-8351},
  \urlprefix\url{https://doi.org/10.1007/s41114-018-0012-9}.

\bibitem[{\citenamefont{Martynov et~al.}(2016)\citenamefont{Martynov, Hall,
  Abbott, Abbott, Abbott, Adams, Adhikari, Anderson, Anderson, Arai
  et~al.}}]{PhysRevD.93.112004}
\bibinfo{author}{\bibfnamefont{D.~V.} \bibnamefont{Martynov}},
  \bibinfo{author}{\bibfnamefont{E.~D.} \bibnamefont{Hall}},
  \bibinfo{author}{\bibfnamefont{B.~P.} \bibnamefont{Abbott}},
  \bibinfo{author}{\bibfnamefont{R.}~\bibnamefont{Abbott}},
  \bibinfo{author}{\bibfnamefont{T.~D.} \bibnamefont{Abbott}},
  \bibinfo{author}{\bibfnamefont{C.}~\bibnamefont{Adams}},
  \bibinfo{author}{\bibfnamefont{R.~X.} \bibnamefont{Adhikari}},
  \bibinfo{author}{\bibfnamefont{R.~A.} \bibnamefont{Anderson}},
  \bibinfo{author}{\bibfnamefont{S.~B.} \bibnamefont{Anderson}},
  \bibinfo{author}{\bibfnamefont{K.}~\bibnamefont{Arai}}, \bibnamefont{et~al.},
  \bibinfo{journal}{Phys. Rev. D} \textbf{\bibinfo{volume}{93}},
  \bibinfo{pages}{112004} (\bibinfo{year}{2016}),
  \urlprefix\url{https://link.aps.org/doi/10.1103/PhysRevD.93.112004}.

\bibitem[{\citenamefont{Aasi et~al.}(2013{\natexlab{a}})\citenamefont{Aasi,
  Abadie, Abbott, Abbott, Abbott, Abernathy, Adams, Adams, Addesso, Adhikari
  et~al.}}]{Squeeze}
\bibinfo{author}{\bibfnamefont{J.}~\bibnamefont{Aasi}},
  \bibinfo{author}{\bibfnamefont{J.}~\bibnamefont{Abadie}},
  \bibinfo{author}{\bibfnamefont{B.~P.} \bibnamefont{Abbott}},
  \bibinfo{author}{\bibfnamefont{R.}~\bibnamefont{Abbott}},
  \bibinfo{author}{\bibfnamefont{T.~D.} \bibnamefont{Abbott}},
  \bibinfo{author}{\bibfnamefont{M.~R.} \bibnamefont{Abernathy}},
  \bibinfo{author}{\bibfnamefont{C.}~\bibnamefont{Adams}},
  \bibinfo{author}{\bibfnamefont{T.}~\bibnamefont{Adams}},
  \bibinfo{author}{\bibfnamefont{P.}~\bibnamefont{Addesso}},
  \bibinfo{author}{\bibfnamefont{R.~X.} \bibnamefont{Adhikari}},
  \bibnamefont{et~al.}, \bibinfo{journal}{Nature Photonics}
  \textbf{\bibinfo{volume}{7}}, \bibinfo{pages}{613}
  (\bibinfo{year}{2013}{\natexlab{a}}).

\bibitem[{\citenamefont{Holland-Ashford
  et~al.}(2017)\citenamefont{Holland-Ashford, Lopez, Auchettl, Temim, and
  Ramirez-Ruiz}}]{Holland-Ashford:2017tqz}
\bibinfo{author}{\bibfnamefont{T.}~\bibnamefont{Holland-Ashford}},
  \bibinfo{author}{\bibfnamefont{L.~A.} \bibnamefont{Lopez}},
  \bibinfo{author}{\bibfnamefont{K.}~\bibnamefont{Auchettl}},
  \bibinfo{author}{\bibfnamefont{T.}~\bibnamefont{Temim}}, \bibnamefont{and}
  \bibinfo{author}{\bibfnamefont{E.}~\bibnamefont{Ramirez-Ruiz}},
  \bibinfo{journal}{Astrophys. J.} \textbf{\bibinfo{volume}{844}},
  \bibinfo{pages}{84} (\bibinfo{year}{2017}), \eprint{1705.08454}.

\bibitem[{\citenamefont{Abadie et~al.}(2011)\citenamefont{Abadie, Abbott,
  Abbott, Abernathy, Accadia, Acernese, Adams, Adhikari, Ajith, Allen
  et~al.}}]{PhysRevLett.107.271102}
\bibinfo{author}{\bibfnamefont{J.}~\bibnamefont{Abadie}},
  \bibinfo{author}{\bibfnamefont{B.~P.} \bibnamefont{Abbott}},
  \bibinfo{author}{\bibfnamefont{R.}~\bibnamefont{Abbott}},
  \bibinfo{author}{\bibfnamefont{M.}~\bibnamefont{Abernathy}},
  \bibinfo{author}{\bibfnamefont{T.}~\bibnamefont{Accadia}},
  \bibinfo{author}{\bibfnamefont{F.}~\bibnamefont{Acernese}},
  \bibinfo{author}{\bibfnamefont{C.}~\bibnamefont{Adams}},
  \bibinfo{author}{\bibfnamefont{R.}~\bibnamefont{Adhikari}},
  \bibinfo{author}{\bibfnamefont{P.}~\bibnamefont{Ajith}},
  \bibinfo{author}{\bibfnamefont{B.}~\bibnamefont{Allen}}, \bibnamefont{et~al.}
  (\bibinfo{collaboration}{LIGO Scientific Collaboration and Virgo
  Collaboration}), \bibinfo{journal}{Phys. Rev. Lett.}
  \textbf{\bibinfo{volume}{107}}, \bibinfo{pages}{271102}
  (\bibinfo{year}{2011}),
  \urlprefix\url{https://link.aps.org/doi/10.1103/PhysRevLett.107.271102}.

\bibitem[{\citenamefont{{Arnett}}(1994)}]{1994ApJ...427..932A}
\bibinfo{author}{\bibfnamefont{D.}~\bibnamefont{{Arnett}}},
  \bibinfo{journal}{\apj} \textbf{\bibinfo{volume}{427}}, \bibinfo{pages}{932}
  (\bibinfo{year}{1994}).

\bibitem[{\citenamefont{{Arnett} and {Meakin}}(2011)}]{2011ApJ...733...78A}
\bibinfo{author}{\bibfnamefont{W.~D.} \bibnamefont{{Arnett}}} \bibnamefont{and}
  \bibinfo{author}{\bibfnamefont{C.}~\bibnamefont{{Meakin}}},
  \bibinfo{journal}{\apj} \textbf{\bibinfo{volume}{733}}, \bibinfo{eid}{78}
  (\bibinfo{year}{2011}), \eprint{1101.5646}.

\bibitem[{\citenamefont{Couch et~al.}(2015)\citenamefont{Couch, Chatzopoulos,
  Arnett, and Timmes}}]{2041-8205-808-1-L21}
\bibinfo{author}{\bibfnamefont{S.~M.} \bibnamefont{Couch}},
  \bibinfo{author}{\bibfnamefont{E.}~\bibnamefont{Chatzopoulos}},
  \bibinfo{author}{\bibfnamefont{W.~D.} \bibnamefont{Arnett}},
  \bibnamefont{and} \bibinfo{author}{\bibfnamefont{F.~X.}
  \bibnamefont{Timmes}}, \bibinfo{journal}{The Astrophysical Journal Letters}
  \textbf{\bibinfo{volume}{808}}, \bibinfo{pages}{L21} (\bibinfo{year}{2015}),
  \urlprefix\url{http://stacks.iop.org/2041-8205/808/i=1/a=L21}.

\bibitem[{\citenamefont{Wongwathanarat
  et~al.}(2010)\citenamefont{Wongwathanarat, Janka, and
  Müller}}]{2041-8205-725-1-L106}
\bibinfo{author}{\bibfnamefont{A.}~\bibnamefont{Wongwathanarat}},
  \bibinfo{author}{\bibfnamefont{H.-T.} \bibnamefont{Janka}}, \bibnamefont{and}
  \bibinfo{author}{\bibfnamefont{E.}~\bibnamefont{Müller}},
  \bibinfo{journal}{The Astrophysical Journal Letters}
  \textbf{\bibinfo{volume}{725}}, \bibinfo{pages}{L106} (\bibinfo{year}{2010}),
  \urlprefix\url{http://stacks.iop.org/2041-8205/725/i=1/a=L106}.

\bibitem[{\citenamefont{{Wongwathanarat, A.}
  et~al.}(2013)\citenamefont{{Wongwathanarat, A.}, {Janka, H.-Th.}, and
  {M\"uller, E.}}}]{refId0}
\bibinfo{author}{\bibnamefont{{Wongwathanarat, A.}}},
  \bibinfo{author}{\bibnamefont{{Janka, H.-Th.}}}, \bibnamefont{and}
  \bibinfo{author}{\bibnamefont{{M\"uller, E.}}}, \bibinfo{journal}{A\&A}
  \textbf{\bibinfo{volume}{552}}, \bibinfo{pages}{A126} (\bibinfo{year}{2013}),
  \urlprefix\url{https://doi.org/10.1051/0004-6361/201220636}.

\bibitem[{\citenamefont{Palmer et~al.}(2005)\citenamefont{Palmer, Barthelmy,
  Gehrels, Kippen, Cayton, Kouveliotou, Eichler, Wijers, Woods, Granot
  et~al.}}]{MagnetarGF}
\bibinfo{author}{\bibfnamefont{D.~M.} \bibnamefont{Palmer}},
  \bibinfo{author}{\bibfnamefont{S.}~\bibnamefont{Barthelmy}},
  \bibinfo{author}{\bibfnamefont{N.}~\bibnamefont{Gehrels}},
  \bibinfo{author}{\bibfnamefont{R.~M.} \bibnamefont{Kippen}},
  \bibinfo{author}{\bibfnamefont{T.}~\bibnamefont{Cayton}},
  \bibinfo{author}{\bibfnamefont{C.}~\bibnamefont{Kouveliotou}},
  \bibinfo{author}{\bibfnamefont{D.}~\bibnamefont{Eichler}},
  \bibinfo{author}{\bibfnamefont{R.~A. M.~J.} \bibnamefont{Wijers}},
  \bibinfo{author}{\bibfnamefont{P.~M.} \bibnamefont{Woods}},
  \bibinfo{author}{\bibfnamefont{J.}~\bibnamefont{Granot}},
  \bibnamefont{et~al.}, \bibinfo{journal}{Nature}
  \textbf{\bibinfo{volume}{434}}, \bibinfo{pages}{1107 EP }
  (\bibinfo{year}{2005}), \urlprefix\url{https://doi.org/10.1038/nature03525}.

\bibitem[{\citenamefont{Zink et~al.}(2012)\citenamefont{Zink, Lasky, and
  Kokkotas}}]{PhysRevD.85.024030}
\bibinfo{author}{\bibfnamefont{B.}~\bibnamefont{Zink}},
  \bibinfo{author}{\bibfnamefont{P.~D.} \bibnamefont{Lasky}}, \bibnamefont{and}
  \bibinfo{author}{\bibfnamefont{K.~D.} \bibnamefont{Kokkotas}},
  \bibinfo{journal}{Phys. Rev. D} \textbf{\bibinfo{volume}{85}},
  \bibinfo{pages}{024030} (\bibinfo{year}{2012}),
  \urlprefix\url{https://link.aps.org/doi/10.1103/PhysRevD.85.024030}.

\bibitem[{\citenamefont{Abbott et~al.}(2009)\citenamefont{Abbott, Abbott,
  Adhikari, Ajith, Allen, Allen, Amin, Anderson, Anderson, Arain
  et~al.}}]{LIGO_SGR1806}
\bibinfo{author}{\bibfnamefont{B.~P.} \bibnamefont{Abbott}},
  \bibinfo{author}{\bibfnamefont{R.}~\bibnamefont{Abbott}},
  \bibinfo{author}{\bibfnamefont{R.}~\bibnamefont{Adhikari}},
  \bibinfo{author}{\bibfnamefont{P.}~\bibnamefont{Ajith}},
  \bibinfo{author}{\bibfnamefont{B.}~\bibnamefont{Allen}},
  \bibinfo{author}{\bibfnamefont{G.}~\bibnamefont{Allen}},
  \bibinfo{author}{\bibfnamefont{R.~S.} \bibnamefont{Amin}},
  \bibinfo{author}{\bibfnamefont{S.~B.} \bibnamefont{Anderson}},
  \bibinfo{author}{\bibfnamefont{W.~G.} \bibnamefont{Anderson}},
  \bibinfo{author}{\bibfnamefont{M.~A.} \bibnamefont{Arain}},
  \bibnamefont{et~al.}, \bibinfo{journal}{The Astrophysical Journal Letters}
  \textbf{\bibinfo{volume}{701}}, \bibinfo{pages}{L68} (\bibinfo{year}{2009}),
  \urlprefix\url{http://stacks.iop.org/1538-4357/701/i=2/a=L68}.

\bibitem[{\citenamefont{Abramowicz et~al.}(2009)\citenamefont{Abramowicz,
  Becker, Biermann, Garzilli, Johansson, and Qian}}]{0004-637X-705-1-659}
\bibinfo{author}{\bibfnamefont{M.~A.} \bibnamefont{Abramowicz}},
  \bibinfo{author}{\bibfnamefont{J.~K.} \bibnamefont{Becker}},
  \bibinfo{author}{\bibfnamefont{P.~L.} \bibnamefont{Biermann}},
  \bibinfo{author}{\bibfnamefont{A.}~\bibnamefont{Garzilli}},
  \bibinfo{author}{\bibfnamefont{F.}~\bibnamefont{Johansson}},
  \bibnamefont{and} \bibinfo{author}{\bibfnamefont{L.}~\bibnamefont{Qian}},
  \bibinfo{journal}{The Astrophysical Journal} \textbf{\bibinfo{volume}{705}},
  \bibinfo{pages}{659} (\bibinfo{year}{2009}),
  \urlprefix\url{http://stacks.iop.org/0004-637X/705/i=1/a=659}.

\bibitem[{\citenamefont{Abramowicz et~al.}(2018)\citenamefont{Abramowicz,
  Bejger, and Wielgus}}]{0004-637X-868-1-17}
\bibinfo{author}{\bibfnamefont{M.~A.} \bibnamefont{Abramowicz}},
  \bibinfo{author}{\bibfnamefont{M.}~\bibnamefont{Bejger}}, \bibnamefont{and}
  \bibinfo{author}{\bibfnamefont{M.}~\bibnamefont{Wielgus}},
  \bibinfo{journal}{The Astrophysical Journal} \textbf{\bibinfo{volume}{868}},
  \bibinfo{pages}{17} (\bibinfo{year}{2018}),
  \urlprefix\url{http://stacks.iop.org/0004-637X/868/i=1/a=17}.

\bibitem[{\citenamefont{Aasi et~al.}(2013{\natexlab{b}})\citenamefont{Aasi,
  Abadie, Abbott, Abbott, Abbott, Abernathy, Accadia, Acernese, Adams, Adams
  et~al.}}]{PhysRevD.88.102002}
\bibinfo{author}{\bibfnamefont{J.}~\bibnamefont{Aasi}},
  \bibinfo{author}{\bibfnamefont{J.}~\bibnamefont{Abadie}},
  \bibinfo{author}{\bibfnamefont{B.~P.} \bibnamefont{Abbott}},
  \bibinfo{author}{\bibfnamefont{R.}~\bibnamefont{Abbott}},
  \bibinfo{author}{\bibfnamefont{T.}~\bibnamefont{Abbott}},
  \bibinfo{author}{\bibfnamefont{M.~R.} \bibnamefont{Abernathy}},
  \bibinfo{author}{\bibfnamefont{T.}~\bibnamefont{Accadia}},
  \bibinfo{author}{\bibfnamefont{F.}~\bibnamefont{Acernese}},
  \bibinfo{author}{\bibfnamefont{C.}~\bibnamefont{Adams}},
  \bibinfo{author}{\bibfnamefont{T.}~\bibnamefont{Adams}}, \bibnamefont{et~al.}
  (\bibinfo{collaboration}{LIGO Scientific Collaboration and Virgo
  Collaboration}), \bibinfo{journal}{Phys. Rev. D}
  \textbf{\bibinfo{volume}{88}}, \bibinfo{pages}{102002}
  (\bibinfo{year}{2013}{\natexlab{b}}),
  \urlprefix\url{https://link.aps.org/doi/10.1103/PhysRevD.88.102002}.

\end{thebibliography}


\end{document}